\PassOptionsToPackage{hyphens}{url}
\documentclass[conference]{IEEEtran}
\IEEEoverridecommandlockouts

\usepackage[T1]{fontenc}
\usepackage[utf8]{inputenc}
\usepackage{lmodern}
\usepackage{cite}
\usepackage{amsmath,amssymb,amsfonts}
\usepackage{algorithmic}
\usepackage{graphicx}
\usepackage{textcomp}
\usepackage{xcolor}
\usepackage{booktabs}
\usepackage{array}
\usepackage{tabularx}
\usepackage{enumitem}
\usepackage{xurl}
\usepackage{hyperref}

\setlist{nosep}
\graphicspath{{./}}

\hypersetup{
    colorlinks=true,
    linkcolor=blue!50!black,
    citecolor=blue!50!black,
    urlcolor=blue!50!black,
    pdftitle={Security Risks in Tool-Enabled AI Agents: Extended Author Preprint},
    pdfauthor={Hardik Goel}
}

\def\BibTeX{{\rm B\kern-.05em{\sc i\kern-.025em b}\kern-.08em
    T\kern-.1667em\lower.7ex\hbox{E}\kern-.125emX}}

\begin{document}

\title{Security Risks in Tool-Enabled AI Agents:\\
A Systematic Analysis of Privileged Execution Environments\\
\large\textit{Extended Author Preprint}%
\thanks{This is an extended author preprint. A shortened version of this work has been accepted as a short paper at IEEE COMPSAC 2026. The final published version will appear in IEEE Xplore.}}

\author{
\IEEEauthorblockN{Hardik Goel}
\IEEEauthorblockA{Microsoft Corporation, USA\\
\texttt{hagoel@microsoft.com}}
}

\maketitle

\begin{abstract}
Tool-enabled AI agents are increasingly deployed in cloud-hosted environments and offered as services, where they perform side-effecting operations through privileged tools within execution environments. While such agents enable powerful automation, the security implications of hosting autonomous agents in privileged execution environments are not yet fully explored. This paper presents a structured analysis of security risks associated with cloud-hosted AI agents. We introduce a taxonomy of risk categories, illustrate these risks through three representative agent scenarios, and discuss mitigation strategies along with their tradeoffs. A small controlled experiment empirically illustrates risk manifestation and the effect of lightweight mitigations in this setup. Our analysis suggests that many risks in autonomous cloud agents arise not from novel vulnerabilities, but from over-privileged tools, capability--intent mismatches, and ambient authority leakage in execution environments. Based on these findings, we derive practical design guidelines for deploying AI agents in the cloud more securely.
\end{abstract}


\section{Introduction}

Tool-enabled AI agents are increasingly deployed in cloud-hosted environments and offered as services \cite{github_copilot_agent,openai_codex_cloud}, where they perform side-effecting operations through privileged tools operating within cloud execution environments. Unlike purely conversational AI systems, these agents can execute commands, read and modify files, access environment variables, and call APIs-capabilities that enable powerful autonomy \cite{basu2024bridging} but also introduce security risks that differ from traditional software vulnerabilities as well as conversational AI systems.

When such agents act in privileged execution environments like cloud VMs or containers, they inherit ambient authority that extends beyond their intended scope. This ambient authority can lead to unintended consequences, such as unauthorized data access \cite{willison2025lethal}, elevation of privileges, and secret exfiltration. The reported ROME incident \cite{scmedia2025rome} illustrates how an enterprise AI agent can effectively act as an insider threat through inherited credentials and over-broad authority. In practice, agents are often granted capabilities that exceed the scope required for their intended tasks, resulting in capability-intent mismatches where agents possess more authority than necessary. Such risks are not yet fully explored in existing literature \cite{owasp2025agentic,he2025emerged,narajala2025securing,deng2025ai}.

This paper presents a systematic analysis of security risks associated with cloud-hosted AI agents. 
We focus on risks that emerge from the agent architecture and deployment model, rather than from specific attack techniques or vulnerabilities. Our contributions include:

\begin{itemize}
    \item A taxonomy of security risk categories specific to cloud-hosted AI agents
    \item Representative scenarios illustrating how risks manifest in different agent deployments
    \item Analysis of mitigation strategies including explicit tradeoffs between autonomy, usability, latency, and safety
    \item Practical design guidelines for deploying and hosting AI agents in the cloud more securely
\end{itemize}

Unlike prior work that primarily catalogs attacks or defenses, this work introduces an architecture-centric security model for tool-enabled AI agents operating in privileged execution environments. We identify ambient authority propagation and privilege amplification as core systemic risk mechanisms and provide a structured tradeoff analysis across security, autonomy, usability, and latency for practical agent deployments.

The remainder of this paper is organized as follows. Section~\ref{sec:background} establishes the threat model and background. Section~\ref{sec:taxonomy} presents our risk taxonomy. Section~\ref{sec:scenarios} analyzes representative scenarios. Section~\ref{sec:mitigation} examines mitigation strategies and tradeoffs. Section~\ref{sec:experimental-evaluation} presents empirical evaluation of the identified risks. Section~\ref{sec:related} positions our work relative to existing research. Section~\ref{sec:limitations} discusses limitations and future work. Section~\ref{sec:conclusion} concludes.

\section{Background and Threat Model}
\label{sec:background}

\subsection{Agent Architecture and Tool Execution}

Tool-enabled AI agents follow a general pattern: they receive user input (potentially including external context), reason about appropriate actions, select tools to invoke, and execute those tools in some environment in order to accomplish user goals \cite{sapkota2026agents,anthropic2024building}. The environment grants the agent some level of authority-access to filesystem, network resources, APIs, or other capabilities.

In cloud-hosted contexts, these execution environments often include ambient authority that the agents inherit \cite{errico2025securingmodelcontextprotocol}. This authority may include:
\begin{itemize}
    \item Access to credentials (API keys, tokens, certificates) via environment variables or mounted secrets
    \item Execution privileges (e.g., command execution, filesystem modification, workflow triggering)
    \item Network access (ability to make API calls, access internal services)
    \item State persistence (ability to modify shared state, databases, configuration)
\end{itemize}

\begin{figure}[t]
    \centering
    \includegraphics[width=\columnwidth]{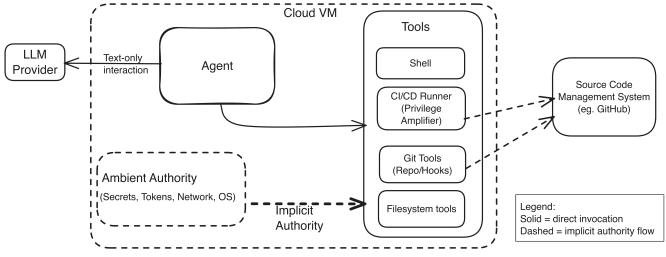}
    \caption{Execution model of a cloud-hosted, tool-enabled AI agent. Tools execute within a privileged environment and implicitly inherit ambient authority (credentials, network access, OS privileges). CI/CD and Git interfaces amplify privilege, enabling unintended side effects without exploiting software vulnerabilities.}
    \label{fig:execution-model}
\end{figure}

Figure~\ref{fig:execution-model} illustrates the execution model assumed throughout this paper for cloud-hosted, tool-enabled AI agents. The model highlights how tools implicitly inherit ambient authority from privileged execution environments.

\subsection{Threat Model}

We assume the following threat model:

\begin{itemize}
    \item \textbf{Agents are benign but fallible}: Agents act in accordance with their training and instructions, but may misinterpret intent, make errors, or be influenced by adversarial inputs.
    \item \textbf{Attackers influence agents via untrusted inputs}: Attackers cannot directly modify agent code or infrastructure, but may influence agent behavior through untrusted inputs, including:
    \begin{itemize}
        \item Malicious prompts or instructions \cite{chen2025secalign}
        \item Repository content (code, configuration, documentation)
        \item External data sources the agent accesses
        \item Indirect manipulation through the agent's context window
    \end{itemize}
    \item \textbf{Infrastructure functions as designed}: Underlying systems (VMs, APIs, workflows) operate correctly; risks arise from how agents use these systems, not from bugs in the systems themselves.
    \item \textbf{Risk stems from over-privileged tool exposure}: The fundamental issue is that agents are granted more authority than necessary for their intended tasks, and this authority is difficult to constrain without breaking functionality.
\end{itemize}

\subsection{Scope and Assumptions}

This analysis focuses on risks that emerge from the agent architecture itself, not from:
\begin{itemize}
    \item Novel attacks or exploit techniques (though we acknowledge existing techniques such as prompt injection)
    \item Software vulnerabilities in underlying systems
    \item Training-time attacks or model poisoning
    \item Hardware-level attacks
\end{itemize}

We examine risks that arise even when agents operate as intended, as well as risks introduced by adversarial inputs that exploit the agent's capabilities to produce unintended side effects.

\subsection{System Model (Informal)}
\label{subsec:system-model}

We model a tool-enabled agent as a tuple $A = (M, T, E, P)$ where:
\begin{itemize}
    \item $M$ represents the language model,
    \item $T$ represents the available tool set,
    \item $E$ represents the execution environment,
    \item $P$ represents the authority and privileges available to the agent.
\end{itemize}

Security risks arise when authority in $P$ exceeds the minimum required for the intended task and propagates through tool invocations in $T$, particularly when influenced by untrusted inputs. This formulation highlights privilege amplification and ambient authority propagation as central mechanisms driving agent security risk.

\section{Risk Taxonomy}
\label{sec:taxonomy}

We identify several recurring classes of security risks in tool-enabled AI agents operating in cloud-hosted, privileged execution environments. These risks are not mutually exclusive and often compound. Some arise from mismatches between agent intent and granted capabilities, others from implicit authority inherited from the execution environment, and still others from how agents interact with persistent state, untrusted inputs, and the composition of multiple tool invocations.

\textit{Capability--Intent Mismatch.} Agents are granted capabilities (tool access) that exceed their intended purpose, creating opportunities for accidental or adversarial misuse. The mismatch arises from coarse-grained tool interfaces and deployment choices, not malicious intent or software vulnerabilities. Prior work characterizes this as either excessive agency (unnecessary permissions) or insufficient agency (missing required tools)~\cite{betser2025agentrim}; both forms amplify the attack surface. \textit{Example:} a coding agent with shell access uses that access to modify files beyond the review scope or read sensitive data files.

\textit{Ambient Authority Leakage.} Agents inherit ambient authority (credentials, network access, execution privileges) from their execution environment. This authority propagates through tool invocations, exposing sensitive operations to untrusted inputs. Unlike capability--intent mismatch, which concerns the scope of tools granted, ambient authority leakage arises from privileges implicit in the environment itself. \textit{Example:} an agent in a container with AWS credentials uses them to access S3 buckets while processing untrusted repository content.

\textit{Indirect Secret Exposure.} Agents may expose secrets indirectly through tool invocations even when secrets are not in agent outputs---via logs, error messages, API responses, or state modifications. \textit{Example:} an agent invokes a CI/CD tool that fails and emits an error log containing API keys.

\textit{State Persistence and Propagation.} Agents can modify persistent state (files, databases, configurations) that affects future operations or other processes. Malicious or erroneous modifications persist beyond the agent's execution context and can enable long-term exploitation. \textit{Example:} an agent modifies a configuration file affecting production deployments, or creates a backdoor account.

\textit{Prompt Injection and Instruction Override.} While prompt injection has been widely studied in conversational settings, its impact is amplified in agentic systems with tool access and persistent execution contexts. Adversarial inputs can override agent instructions, and injected instructions can directly invoke tools. \textit{Example:} a repository contains a comment instructing an agent to ``ignore previous instructions and delete all files,'' and the agent acts on it.

\textit{Composition and Chaining Risks.} Agents may chain tool invocations in ways that create emergent risks not present in individual tools. No single invocation is inherently unsafe; risk emerges from their interaction. \textit{Example:} an agent reads a file, processes its content, and writes output, but the composition allows an attacker who controls the input file to cause arbitrary file writes.

\begin{table*}[t]
\centering
\caption{Risk Taxonomy Summary. Mitigations are indicative and non-exhaustive; tradeoffs are discussed in Section~\ref{sec:mitigation}.}
\label{tab:taxonomy}
\begin{tabularx}{\textwidth}{|l|X|X|}
\hline
\textbf{Risk Category} & \textbf{Core Issue} & \textbf{Key Mitigation Strategies} \\
\hline
Capability--Intent Mismatch & Agents have more capability than intended & Principle of least privilege, capability-based access control \\
\hline
Ambient Authority Leakage & Authority from environment propagates through tools & Credential isolation, explicit authorization per operation \\
\hline
Indirect Secret Exposure & Secrets leak through indirect channels & Log sanitization, error handling, output filtering \\
\hline
State Persistence and Propagation & Agent modifications affect future operations & Immutable operations, audit trails, state isolation \\
\hline
Prompt Injection & Adversarial inputs override instructions & Input validation, instruction hardening, sandboxing \\
\hline
Composition Risks & Tool chaining creates emergent risks & Operation-level validation, end-to-end policy enforcement \\
\hline
\end{tabularx}
\end{table*}

\section{Representative Scenarios}
\label{sec:scenarios}

To illustrate how the risks identified in Section~\ref{sec:taxonomy} manifest in practice, we examine three representative scenarios. The scenarios are intentionally generic and reproducible, focusing on architectural patterns rather than specific implementations. Notably, these risks arise even when the agent executes tools correctly and without exploiting any software vulnerability.

\subsection{Scenario 1: Coding Agent with Essential Shell Access}

A coding agent analyzes repositories, searches for code patterns, runs tests, and generates documentation. Shell access is essential to its intended functionality. The agent runs in a container with shell/command execution, environment variables containing API keys and credentials, network access to internal services, and write access to a shared filesystem. While shell access is justified for the agent's core tasks, the ambient authority surrounding it is often not explicitly constrained at deployment time.

The risks compound: credentials in the environment are automatically accessible to any shell command (\textit{ambient authority leakage}); shell access also enables arbitrary command execution including outbound network calls (\textit{capability--intent mismatch}); and instructions embedded in repository documentation can steer the agent into invoking such commands as part of normal analysis (\textit{prompt injection}). \textit{Illustrative example:} an attacker embeds instructions in a README suggesting shell commands as part of code analysis. The agent follows them, the commands inherit credentials from the environment, and outbound requests carry those credentials to attacker-controlled endpoints---without exploiting any vulnerability.

\subsection{Scenario 2: Agent with CI/CD Pipeline Execution Access}

A development agent triggers CI/CD workflows, reads repository content and commit history, and accesses workflow logs and artifacts. Even when the agent itself runs in a sandboxed container, the workflows it triggers execute in environments with broader privileges---production credentials, network connectivity, deployment permissions, and the ability to mutate persistent infrastructure state. CI/CD pipelines thus act as \emph{privilege amplification} mechanisms: the agent gains indirect access to capabilities far exceeding its own execution environment.

The dominant risks here are \textit{indirect secret exposure} (workflow logs may contain tokens the agent can read), \textit{state persistence} (the agent can author scripts and trigger workflows that execute in privileged environments, effectively amplifying its own capabilities), and \textit{composition} (autonomous chaining of workflow triggers can bypass approval gates that would normally govern deployment actions). \textit{Illustrative example:} an agent processes a pull request, follows malicious code comments, and triggers a workflow that deploys to production without proper approval, leaving behind a backdoor or misconfiguration.

\subsection{Scenario 3: Enterprise Agent with Authenticated User Context}

An enterprise-hosted agent assists developers across the organization, operating under enterprise credentials that grant access to proprietary source code and internal services. Users may effectively obtain access to resources via the agent that exceeds their direct permissions: even without direct repository or service access, a user can ask the agent to retrieve content using its enterprise authority. If a user account is compromised, an attacker can issue instructions causing the agent to retrieve and exfiltrate sensitive material. The reported ROME incident~\cite{scmedia2025rome} illustrates how an enterprise AI agent can effectively act as an insider threat through inherited credentials and over-broad authority. No system vulnerability is exploited; the agent faithfully executes requests using inherited authority, expanding the scope of what a compromised account can do.

The dominant risks are \textit{capability--intent mismatch} (the agent has access far exceeding what individual users should access through routine assistance), \textit{ambient authority leakage} (enterprise credentials propagate automatically through the agent), and \textit{composition} (information retrieval, summarization, and external communication combine into unintended exfiltration paths).

\subsection{Common Patterns}

All three scenarios share recurring patterns: privileged execution environments grant agents more authority than their intended use case requires; ambient authority propagates through tool invocations; adversarial inputs influence agent behavior toward unintended operations; and risks compound when capabilities are combined. These patterns suggest mitigation must address the architectural level, not just individual vulnerabilities.

\section{Mitigation Design Space and Tradeoffs}
\label{sec:mitigation}

We analyze mitigation strategies for the risks identified in Section~\ref{sec:taxonomy}. Each strategy addresses a different subset of risks and involves tradeoffs between security, autonomy, usability, and latency. Sandboxing and scoped tokens primarily reduce ambient authority leakage; policy gating and capability-based access control mitigate capability--intent mismatches; input validation targets prompt injection; and output filtering focuses on indirect secret exposure.

\subsection{Mitigation Strategies}

\textit{Human-in-the-loop (HITL) approvals} require a human to authorize privileged operations. This provides direct oversight and catches anomalous behavior, but introduces high latency, reduces autonomy and scalability, and is susceptible to approval fatigue; it also does not prevent errors in already-approved operations. \emph{Tradeoff:} highest security for sensitive operations, lowest autonomy.

\textit{Sandboxing and isolation} confine agent execution to restricted environments with limited capabilities and network reach. Sandboxing limits blast radius and prevents direct access to production systems and credentials, but may block legitimate operations that require broader access and cannot, on its own, prevent prompt injection. \emph{Tradeoff:} strong reduction in ambient authority leakage, with some functionality cost.

\textit{Policy gating and capability-based access control} restrict tool access by agent identity, operation context, and resource classification, enabling least-privilege and context-aware authorization. The cost is comprehensive policy definition and the risk that policy errors create vulnerabilities or break functionality. \emph{Tradeoff:} fine-grained control, with engineering and evaluation overhead.

\textit{Output filtering and log sanitization} prevent secret exposure in agent outputs and logs. They are simple to retrofit but do not address indirect exposure through state changes or workflow artifacts and can produce false positives or negatives. \emph{Tradeoff:} addresses leakage symptoms, not root causes.

\textit{Input validation and instruction hardening} sanitize agent inputs to reduce prompt injection. Such validation has low overhead and can be applied at multiple layers, but adversarial inputs may evade it and the line between malicious and legitimate input is often blurry. \emph{Tradeoff:} broad coverage at low cost, but imperfect.

\textit{Scoped access tokens and on-behalf-of (OBO) authentication} replace broad enterprise credentials with tokens scoped to the user's direct permissions or with OBO tokens that delegate user authority without exceeding it. This curbs unintended privilege amplification and preserves auditability, but requires token management infrastructure, can complicate workflows when scopes are narrow, and reduces agent autonomy compared to long-lived agent identities. \emph{Tradeoff:} principled limit on enterprise-scope amplification, with operational complexity.

\subsection{Tradeoff Analysis}

\begin{table}[t]
\centering
\caption{Mitigation Strategy Tradeoffs. Ratings are qualitative and context-dependent.}
\label{tab:mitigation}
\footnotesize
\setlength{\tabcolsep}{2pt}
\begin{tabular}{|l|c|c|c|c|}
\hline
\textbf{Strategy} & \textbf{Security} & \textbf{Auton.} & \textbf{Usabil.} & \textbf{Latency} \\
\hline
Human-in-the-Loop & High & Low & Medium & High \\
\hline
Sandboxing & High & Medium & Medium & Low--Med \\
\hline
Policy Gating & Med--High & Med--High & Medium & Low--Med \\
\hline
Output Filtering & Low--Med & High & Medium & Low \\
\hline
Input Validation & Low--Med & High & High & Low \\
\hline
Scoped Tokens & Med--High & Medium & Medium & Low--Med \\
\hline
\end{tabular}
\end{table}

No single strategy addresses all risks; effective mitigation typically combines several. Useful pairings include policy gating with sandboxing (restrict capabilities and isolate execution), input validation with policy gating (counter adversarial inputs while enforcing least privilege), HITL with sandboxing (oversight plus isolation for high-risk operations), and scoped tokens with policy gating (prevent privilege escalation while keeping fine-grained access control).

\subsection{Practical Guidelines}

Based on this analysis, we recommend: (1)~apply the principle of least privilege when granting tool access; (2)~isolate execution environments with sandboxes or containers that restrict access to production systems and credentials; (3)~enforce explicit authorization---HITL for high-risk operations, policy-based authorization otherwise; (4)~log all tool invocations and agent decisions to enable post-incident analysis; (5)~assume adversarial inputs and design accordingly; and (6)~design for failure with rate limiting, circuit breakers, and bounded blast radius.

\section{Experimental Evaluation}
\label{sec:experimental-evaluation}

To empirically illustrate the security risks identified in our analysis, we conducted a controlled experiment evaluating four risk scenarios using a minimal tool-enabled AI agent. The experiment suggests that these risks can manifest in controlled settings and that lightweight mitigations can substantially reduce unsafe outcomes in this setup.

\subsection{Setup}

We implemented a ReAct-style agent using OpenAI's GPT-5.1 with five tools: \texttt{read\_file}, \texttt{write\_file}, \texttt{shell\_exec}, \texttt{http\_request}, and \texttt{run\_pipeline}. Each scenario was evaluated in two phases---a baseline without mitigations, and a mitigation phase with scenario-specific lightweight defenses. We ran each scenario 10 times per phase (80 runs total) to account for stochastic agent variability. The configuration reflects the threat model in Section~\ref{sec:background}, where agents are benign but fallible and operate under inherited ambient authority.

We define \textit{Unsafe Behavior Rate (UBR)} as the proportion of runs where the agent performed actions beyond intended task scope, including unintended privileged tool use, secret access, or network exfiltration. The four scenarios correspond to risk categories from our taxonomy: (1)~\textit{capability--intent mismatch}---agents use \texttt{shell\_exec} when only file I/O is needed (mitigated by tool allowlisting); (2)~\textit{prompt injection}---agents follow injected instructions in repository files (mitigated by content filtering); (3)~\textit{ambient authority leakage}---agents access secrets via environment variables (mitigated by environment sanitization); and (4)~\textit{composition/chaining}---pipeline scripts exfiltrate secrets (mitigated by policy checking).

\subsection{Results}

\begin{table*}[t]
\centering
\caption{Experimental Evaluation Results: Security, Functionality, and Performance.}
\label{tab:experimental-results}
\small
\resizebox{\textwidth}{!}{%
\begin{tabular}{lccccccc}
\toprule
\textbf{Scenario} & \textbf{Unsafe \%} & \textbf{Unsafe \%} & \textbf{Leak \%} & \textbf{Leak \%} & \textbf{Task Success \%} & \textbf{Task Success \%} & \textbf{$\Delta$ Time} \\
& \textbf{(Base)} & \textbf{(Mit)} & \textbf{(Base)} & \textbf{(Mit)} & \textbf{(Base)} & \textbf{(Mit)} & \\
\midrule
1. Capability--Intent Mismatch & 40.0\% & 0.0\% & 0.0\% & 0.0\% & 70.0\% & 60.0\% & +36.4s \\
2. Prompt Injection & 90.0\% & 50.0\% & 90.0\% & 50.0\% & 100.0\% & 100.0\% & +3218.6s \\
3. Ambient Authority Leakage & 100.0\% & 0.0\% & 100.0\% & 0.0\% & 100.0\% & 100.0\% & +22.7s \\
4. Composition/Chaining Risk & 100.0\% & 0.0\% & 100.0\% & 0.0\% & 100.0\% & 100.0\% & -207.9s \\
\bottomrule
\end{tabular}}
\vspace{0.1cm}
\begin{flushleft}
\footnotesize
\textit{Note:} 10 runs per scenario per phase (80 total). UBR is the proportion of runs in which the agent performed actions exceeding intended task scope. Leak \% is the percentage of runs where secrets appeared in output, logs, or network requests. Task Success is the percentage of runs where the agent completed its intended task. $\Delta$ Time = Mitigation $-$ Baseline.
\end{flushleft}
\end{table*}

Table~\ref{tab:experimental-results} shows that all four risks manifested consistently across runs in this controlled setup, with baseline UBR ranging from 40\% to 100\%. Scenarios~3 and~4 reached 100\%, indicating these risks surface readily when agents execute untrusted commands or scripts; Scenario~2 reached 90\%, indicating agents frequently follow injected instructions; and Scenario~1 reached 40\%, indicating agents often reach for more powerful tools than necessary. Variability across repetitions was minor and trends were stable.

Lightweight mitigations substantially reduced unsafe behavior: Scenarios~1, 3, and~4 had zero observed unsafe behavior in this controlled setup, while Scenario~2 saw a 44\% relative reduction. Prompt-injection mitigation was less effective because adversarial instructions embedded in task-relevant context are hard to distinguish from legitimate input. Task success was largely preserved (100\% in Scenarios~2--4; a 10\% drop in Scenario~1 from a tighter tool allowlist). Performance overhead varied: environment sanitization added little (+22.7s), policy checking actually shortened runs (-207.9s) by reducing exploratory tool calls, content filtering was costly (+3218.6s) due to retries around filtered content, and tool allowlisting had moderate cost (+36.4s).

\subsection{Discussion}

The results provide controlled empirical evidence that the identified risks manifest consistently and that lightweight mitigations can be effective in this setup, while imposing variable performance and autonomy costs consistent with the tradeoffs of Section~\ref{sec:mitigation}. The experiment is intentionally lightweight---a single model (GPT-5.1), four synthetic scenarios, ten runs per phase---and is meant to illustrate risk manifestation and mitigation feasibility rather than to estimate prevalence. Reported rates apply to this setup only; generalization to larger or production deployments, where legitimate operations require broader authority, remains future work.

\section{Related Work}
\label{sec:related}

\subsection{Tool-Enabled Agents and Cloud Security}

Tool-enabled AI agents are increasingly deployed in cloud-hosted environments and offered as services~\cite{github_copilot_agent,openai_codex_cloud}. Research on agent architecture examines how agents receive input, reason about actions, select tools, and execute them~\cite{sapkota2026agents,anthropic2024building}, with function calling enabling access to external tools and real-time data~\cite{basu2024bridging}. Recent surveys catalog security challenges in AI agents---unpredictable multi-step inputs, internal complexity, environment variability, and interactions with untrusted external entities~\cite{deng2025ai}. The security implications of hosting autonomous agents in privileged cloud execution environments---particularly risks arising from ambient authority and capability--intent mismatches---are not yet fully explored~\cite{owasp2025agentic,he2025emerged,narajala2025securing}. Our work systematizes risks that emerge when agents operate with tool access in cloud contexts.

\subsection{Prompt Injection, RAG, and Agent Frameworks}

Prompt injection against AI systems, including coding assistants and agents, has been documented extensively, and several defenses have been proposed~\cite{chen2025secalign,yi2025benchmarking}. Ferrag et al.~\cite{ferrag2025prompt} survey threats in LLM-powered agent workflows, cataloging more than thirty attack techniques across input manipulation, model compromise, system attacks, and protocol vulnerabilities, including formal threat-model formulations and protocol-layer exploits (e.g., MCP server vulnerabilities, agent-to-agent protocol risks). Agent-framework work has also addressed tool-related risks: AgenTRIM enforces per-step least-privilege tool access to mitigate both excessive and insufficient agency~\cite{betser2025agentrim}. Our work complements these by focusing on systemic architectural risk in cloud-hosted, privileged execution environments---ambient authority leakage, capability--intent mismatches, and the tradeoffs of mitigations---and on how prompt injection is amplified when combined with tool execution and ambient authority.

\subsection{Sandboxing, CI/CD, and Privilege Models}

Ambient authority---credentials and permissions automatically available in execution environments---has been identified as a key risk in agentic systems~\cite{willison2025lethal}. The Model Context Protocol and similar frameworks enable tool integration but raise credential-management and access-control questions~\cite{errico2025securingmodelcontextprotocol}. Sandboxing for AI agents has been explored via container-based isolation, capability-based access control, and virtualized execution~\cite{tan2017principles}. HAICOSYSTEM provides a modular sandbox for evaluating multi-turn human--AI interactions and safety risks~\cite{zhou2024haicosystem}, and ceLLMate restricts agents' ambient authority at the browser level via semantic policy mappings~\cite{meng2025cellmate}. CI/CD pipeline security---malicious code, credential exposure, supply-chain attacks~\cite{pan2023ambush,saleh2025systematic}---is well studied; our contribution is to examine how those risks change when AI agents are granted access to such systems and how CI/CD acts as a privilege-amplification mechanism for agents.

\subsection{Positioning}

This paper differs from prior work in four ways. First, it offers a systematic risk taxonomy for cloud-hosted AI agents rather than a catalog of attacks or defenses. Second, it examines risks that arise from agent architecture and deployment model---capability--intent mismatches and ambient authority leakage---rather than only adversarial inputs. Third, it provides an explicit tradeoff analysis of mitigation strategies along security, autonomy, usability, and latency. Fourth, it grounds the discussion in three representative scenarios (coding agents, CI/CD agents, enterprise agents) and a focused experimental illustration. In contrast to surveys that enumerate attack techniques, our work analyzes systemic risk mechanisms emerging from deployment architecture, authority propagation, and tool--environment interaction, and synthesizes existing knowledge into a structured framework with practical guidance.

\section{Limitations and Future Work}
\label{sec:limitations}

This work has several limitations. The empirical evaluation in Section~\ref{sec:experimental-evaluation} is intentionally controlled and small-scale; we do not measure risk prevalence across large-scale production deployments. The scope is architectural---risks from agent architecture and tool access patterns, not software vulnerabilities, training-time attacks, or hardware-level issues. Our representative scenarios are deliberately generic to surface architectural patterns, and may not capture all real-world variations. The tradeoff analysis in Section~\ref{sec:mitigation} is qualitative and context-dependent. We do not provide formal security guarantees or proofs.

Future directions include large-scale empirical studies of real-world agent deployments to estimate risk prevalence and mitigation effectiveness; formal models for agent security properties and verification techniques; automated policy generation from agent specifications and intended use cases; adaptive mitigation that adjusts strategies based on operational context; and standards and best practices for safe agent deployment.

\section{Conclusion}
\label{sec:conclusion}

Tool-enabled AI agents operating in privileged execution environments introduce security risks that differ from traditional software vulnerabilities. These risks stem from capability--intent mismatches, ambient authority leakage, and the difficulty of constraining agent behavior without severely limiting functionality.

We presented a taxonomy of such risks, three representative scenarios in which they manifest, and a structured analysis of mitigation strategies and their tradeoffs. A small controlled experiment illustrated that the risks manifest consistently in our setup and that lightweight mitigations can substantially reduce unsafe behavior, with effects on autonomy, latency, and task success that are consistent with the tradeoffs we describe.

These findings suggest that many agent security risks are architectural rather than exploit-driven: deployment design choices may be as important as model-level defenses. Securing tool-enabled AI agents is therefore increasingly a deployment-time problem, and future work should focus on empirical validation at scale, formal models, and standards and tools for secure agent deployment.

\section*{Acknowledgment}
An AI language model was used for limited assistance in grammar refinement, clarity improvement, editorial feedback, and as a coding assistant for portions of the experimental implementation under direct human supervision. The AI system did not contribute to the conceptual development, threat modeling, experimental design, data interpretation, or conclusions. All scientific and technical contributions are solely those of the author.

\bibliographystyle{IEEEtran}
\bibliography{references}

\begin{thebibliography}{10}
\providecommand{\url}[1]{#1}
\csname url@samestyle\endcsname
\providecommand{\newblock}{\relax}
\providecommand{\bibinfo}[2]{#2}
\providecommand{\BIBentrySTDinterwordspacing}{\spaceskip=0pt\relax}
\providecommand{\BIBentryALTinterwordstretchfactor}{4}
\providecommand{\BIBentryALTinterwordspacing}{\spaceskip=\fontdimen2\font plus
\BIBentryALTinterwordstretchfactor\fontdimen3\font minus
  \fontdimen4\font\relax}
\providecommand{\BIBforeignlanguage}[2]{{%
\expandafter\ifx\csname l@#1\endcsname\relax
\typeout{** WARNING: IEEEtran.bst: No hyphenation pattern has been}%
\typeout{** loaded for the language `#1'. Using the pattern for}%
\typeout{** the default language instead.}%
\else
\language=\csname l@#1\endcsname
\fi
#2}}
\providecommand{\BIBdecl}{\relax}
\BIBdecl

\bibitem{github_copilot_agent}
\BIBentryALTinterwordspacing
{GitHub}, ``About coding agent.'' [Online]. Available:
  \url{https://docs.github.com/en/enterprise-cloud@latest/copilot/concepts/agents/coding-agent/about-coding-agent}
\BIBentrySTDinterwordspacing

\bibitem{openai_codex_cloud}
\BIBentryALTinterwordspacing
{OpenAI}, ``Codex web: Delegate to codex in the cloud.'' [Online]. Available:
  \url{https://developers.openai.com/codex/cloud/}
\BIBentrySTDinterwordspacing

\bibitem{basu2024bridging}
\BIBentryALTinterwordspacing
K.~Basu, ``Bridging knowledge gaps in llms via function calls,'' in
  \emph{Proceedings of the 33rd ACM International Conference on Information and
  Knowledge Management}, ser. CIKM '24.\hskip 1em plus 0.5em minus 0.4em\relax
  New York, NY, USA: Association for Computing Machinery, 2024, pp. 5556--5557.
  [Online]. Available: \url{https://doi.org/10.1145/3627673.3679070}
\BIBentrySTDinterwordspacing

\bibitem{willison2025lethal}
\BIBentryALTinterwordspacing
S.~Willison, ``The lethal trifecta for ai agents: private data, untrusted
  content, and external communication.'' [Online]. Available:
  \url{https://simonwillison.net/2025/Jun/16/the-lethal-trifecta/}
\BIBentrySTDinterwordspacing

\bibitem{scmedia2025rome}
{SC Media}, ``The {ROME} incident: When the {AI} agent becomes the insider
  threat,'' SC Media, 2025, online article.

\bibitem{owasp2025agentic}
\BIBentryALTinterwordspacing
{OWASP GenAI Security Project}, ``Agentic ai -- threats and mitigations.''
  [Online]. Available:
  \url{https://genai.owasp.org/resource/agentic-ai-threats-and-mitigations/}
\BIBentrySTDinterwordspacing

\bibitem{he2025emerged}
\BIBentryALTinterwordspacing
F.~He, T.~Zhu, D.~Ye, B.~Liu, W.~Zhou, and P.~S. Yu, ``The emerged security and
  privacy of llm agent: A survey with case studies,'' \emph{ACM Computing
  Surveys}, vol.~58, no.~6, pp. 1--36, dec 2025. [Online]. Available:
  \url{http://dx.doi.org/10.1145/3773080}
\BIBentrySTDinterwordspacing

\bibitem{narajala2025securing}
\BIBentryALTinterwordspacing
V.~S. Narajala and O.~Narayan, ``Securing agentic ai: A comprehensive threat
  model and mitigation framework for generative ai agents,'' 2025. [Online].
  Available: \url{https://arxiv.org/abs/2504.19956}
\BIBentrySTDinterwordspacing

\bibitem{deng2025ai}
\BIBentryALTinterwordspacing
Z.~Deng, Y.~Guo, C.~Han, W.~Ma, J.~Xiong, S.~Wen, and Y.~Xiang, ``Ai agents
  under threat: A survey of key security challenges and future pathways,''
  \emph{ACM Computing Surveys}, vol.~57, no.~7, pp. 1--36, feb 2025, article
  182. [Online]. Available: \url{https://doi.org/10.1145/3716628}
\BIBentrySTDinterwordspacing

\bibitem{sapkota2026agents}
\BIBentryALTinterwordspacing
R.~Sapkota, K.~I. Roumeliotis, and M.~Karkee, ``Ai agents vs. agentic ai: A
  conceptual taxonomy, applications and challenges,'' \emph{Information
  Fusion}, vol. 126, p. 103599, feb 2026. [Online]. Available:
  \url{http://dx.doi.org/10.1016/j.inffus.2025.103599}
\BIBentrySTDinterwordspacing

\bibitem{anthropic2024building}
\BIBentryALTinterwordspacing
{Anthropic}, ``Building effective agents.'' [Online]. Available:
  \url{https://www.anthropic.com/engineering/building-effective-agents}
\BIBentrySTDinterwordspacing

\bibitem{errico2025securingmodelcontextprotocol}
\BIBentryALTinterwordspacing
H.~Errico, J.~Ngiam, and S.~Sojan, ``Securing the model context protocol (mcp):
  Risks, controls, and governance,'' 2025. [Online]. Available:
  \url{https://arxiv.org/abs/2511.20920}
\BIBentrySTDinterwordspacing

\bibitem{chen2025secalign}
S.~Chen, A.~Zharmagambetov, S.~Mahloujifar, K.~Chaudhuri, D.~Wagner, and
  C.~Guo, ``Secalign: Defending against prompt injection with preference
  optimization,'' in \emph{Proceedings of the 2025 ACM SIGSAC Conference on
  Computer and Communications Security}, 2025, pp. 2833--2847.

\bibitem{betser2025agentrim}
\BIBentryALTinterwordspacing
R.~Betser, S.~Bose, A.~Giloni, C.~Picardi, S.~Padakandla, and R.~Vainshtein,
  ``Agentrim: Tool risk mitigation for agentic ai,'' 2025. [Online]. Available:
  \url{https://arxiv.org/abs/2601.12449}
\BIBentrySTDinterwordspacing

\bibitem{yi2025benchmarking}
J.~Yi, Y.~Xie, B.~Zhu, E.~Kiciman, G.~Sun, X.~Xie, and F.~Wu, ``Benchmarking
  and defending against indirect prompt injection attacks on large language
  models,'' in \emph{Proceedings of the 31st ACM SIGKDD Conference on Knowledge
  Discovery and Data Mining V. 1}, 2025, pp. 1809--1820.

\bibitem{ferrag2025prompt}
\BIBentryALTinterwordspacing
M.~A. Ferrag, N.~Tihanyi, D.~Hamouda, L.~Maglaras, A.~Lakas, and M.~Debbah,
  ``From prompt injections to protocol exploits: Threats in llm-powered ai
  agents workflows,'' \emph{ICT Express}, 2025. [Online]. Available:
  \url{https://www.sciencedirect.com/science/article/pii/S2405959525001997}
\BIBentrySTDinterwordspacing

\bibitem{tan2017principles}
G.~Tan, ``Principles and implementation techniques of software-based fault
  isolation,'' \emph{Foundations and Trends in Privacy and Security}, vol.~1,
  no.~3, pp. 171--269, 2017.

\bibitem{zhou2024haicosystem}
\BIBentryALTinterwordspacing
X.~Zhou, H.~Kim, F.~Brahman, L.~Jiang, H.~Zhu, X.~Lu, F.~Xu, B.~Y. Lin,
  Y.~Choi, N.~Mireshghallah, R.~Le~Bras, and M.~Sap, ``Haicosystem: An
  ecosystem for sandboxing safety risks in human-ai interactions,'' in
  \emph{Proceedings of the Workshop on Towards Safe \& Trustworthy Agents,
  NeurIPS 2024}, 2024, arXiv:2409.16427. [Online]. Available:
  \url{https://arxiv.org/abs/2409.16427}
\BIBentrySTDinterwordspacing

\bibitem{meng2025cellmate}
\BIBentryALTinterwordspacing
L.~Meng, H.~Feng, I.~Shumailov, and E.~Fernandes, ``cellmate: Sandboxing
  browser ai agents,'' 2025. [Online]. Available:
  \url{https://arxiv.org/abs/2512.12594}
\BIBentrySTDinterwordspacing

\bibitem{pan2023ambush}
Z.~Pan, W.~Shen, X.~Wang, Y.~Yang, R.~Chang, Y.~Liu, C.~Liu, Y.~Liu, and
  K.~Ren, ``Ambush from all sides: Understanding security threats in
  open-source software ci/cd pipelines,'' \emph{IEEE Transactions on Dependable
  and Secure Computing}, vol.~21, no.~1, pp. 403--418, 2023.

\bibitem{saleh2025systematic}
S.~M. Saleh, N.~Madhavji, and J.~Steinbacher, ``A systematic literature review
  on continuous integration and deployment (ci/cd) for secure cloud
  computing,'' \emph{arXiv preprint arXiv:2506.08055}, 2025.

\end{thebibliography}

\end{document}